\documentclass{article}  
\usepackage{spadre2008}
\usepackage{graphicx}
\frompage{000} \topage{000}                                              
\def\rightmark{Searching for Jets in Heavy Ion Collisions}
\def\leftmark{S. Salur}
 
\title{Searching for Jets in Heavy Ion Collisions} 
\authors{
{Sevil Salur$^1$}\\[2.812mm]
{\normalsize
\hspace*{-8pt}$^1$ Lawrence Berkeley National Laboratory, 1 Cyclotron Dr., MS-70R0319 \\ 
Berkeley, 94720 CA USA\\[0.2ex] 
}
}
 
\abstract{Jet quenching measurements using leading particles and their 
correlations suffer from known biases, which can be removed via 
direct reconstruction of jets in central heavy ion collisions. 
In this talk, we discuss several modern jet reconstruction algorithms and background subtraction techniques that are appropriate to heavy ion collisions. 
}

 
\begin{document}
 
\maketitle
\setcounter{page}{1}

\section{Introduction}\label{intro}

In high energy collisions, during the hadronization process, colored partons from the hard scatter via soft quark and gluon radiation form a ``spray'' of colorless hadrons also known as ``jets''.   Jets are expected to reflect kinematics and topology of partons and they are the experimental signatures of quarks and gluons \cite{drell,cabibbo,bjorken,weinberg}. The final states of hadrons can be grouped into jets via various algorithms.  All the algorithms that are appropriate for hadronic collisions can be used to crosscheck and estimate the systematics of the measurement of jets and the fragmentation functions. At Tevatron, the inclusive jet cross section is measured for 20 orders of magnitude via sequential and cone algorithms and these measurements are consistent with the NLO pQCD prediction from CTEQ 6.1 \cite{cdfall,cdf}. The first direct measurement of inclusive mid-rapidity jets at RHIC are performed in polarized $p+p$ collisions at  $\sqrt s=200$ GeV with STAR experiment \cite{starpp}. For this measurement, the neutral particles are measured with the barrel calorimeter of STAR with full azimuthal ($0< \phi < 2 \pi $) and partial rapidity  ($0.2< \eta < 0.8$) coverage. The neutral particles together with the charged particles that are identified with the time projection chamber ($0 < \phi < 2 \pi $, $| \eta | < 1.3$) are combined into jets with a mid-point jet cone algorithm where split-merge step is required. This inclusive jet cross section measurement from RHIC at $\sqrt s=200$ GeV polarized $p+p$ collisions also agrees well with the NLO pQCD.

Jets can be used as direct probes of partonic phase in the heavy ion collisions at the RHIC and soon at the LHC \cite{starpt,phenixpt}.  Modified leading logarithmic approximation (MLLA), which provides a good description of vacuum fragmentation, introduces medium effects at parton splitting \cite{borvi}. In this approximation, fragmentation of jets is strongly modified for hadrons with transverse momentum of $p_{T}=1-5$ GeV. The study of fragmentation of jets to probe the high density medium created at heavy ion collisions is similar to the study of deep inelastic scattering experiments as a probe of spatial structure and the forces introduced.  During the last 8 years of RHIC operations, instead of direct reconstruction of jets due to the challenges in understanding the underlying high multiplicity events, measurements of high $p_{T}$ hadron suppression are studied via di-hadron correlations and nuclear modification factors $(R_{AA})$ \cite{starpt,phenixpt}.  While these partonic energy loss measurements are consistent with pQCD based energy loss through medium induced gluon radiation calculations,  surface emission bias limits sensitivity to essential $\hat q $ measurement and they can only provide a lower bound to the initial color charge density \cite{eskola}.  Full unbiased jet reconstruction (i.e., reconstructing partonic kinematics independent of quenched or unquenched fragmentation details)  is required in heavy ion collisions. Jet shapes, fragmentation functions, and energy flow will increase our sensitivity to quenching.  In the following sections, we discuss several modern jet reconstruction algorithms that are appropriate for heavy ion collisions and their background subtraction techniques.

\section{Algorithms Used}\label{techno}

See \cite{blazey,salamtalk,jets} and references therein for an exclusive overview of many traditional jet algorithms used in high energy collisions. Here we discuss only two algorithms; seeded cone and sequential recombination, and their background subtractions.  As in cleaner $e^{+}+e^{-}$ or $p+p$ collisions, in heavy ion collisions the chosen jet reconstruction method should be theoretically and experimentally consistent. For example from the theory side, the algorithm should have low insensitivity to hadronization, radiation, splitting  and be equally well defined at hadron and parton level. The larger  LHC luminosities (20 to 200 collisions in a detector) require that the traditional jet algorithms have to be improved to resolve pile up of events.  The underlying pile up event can be  subtracted with the clustering algorithm and the consideration of the jet area \cite{pileupcacciari,catchment}.  These subtraction  techniques can also be used in a high multiplicity heavy ion environment where the background subtraction is required.  As Weinberg says ``Quark and gluon jets (identified to partons) can be compared to detector jets, if jet algorithms respect collinear and infrared safety.'' \cite{weinberg}. However, the cone algorithms with seeds can be used to estimate the systematics of the background subtraction in heavy ion collisions.  The algorithm should also be detector independent and should allow the combination of particles detected in various detectors such as barrel calorimeter and time projection chamber for neutral and charged particles. As the physical energy measurement degrades the resolution somewhat with the detectors, the choice of algorithm should be aimed to minimize the additional resolution effects.  To handle a large number of events with high multiplicities, the computational efficiency is also required. However, with the improved grid system and continuously increasing technology of computational speed, it is possible to work with slower algorithms \cite{grid}.

\subsection{Cone}\label{cone}

The main principle of the cone algorithm is to combine particles in $\eta - \phi $ space with their neighbors within a cone of radius R ($R=\sqrt{ \Delta \phi ^{2}+ \Delta \eta^{2} }$).  Variation of the algorithms such as seeds (requirement of minimum energy), splitting, merging, iteration and so on exists to optimize the search and effectiveness of jet finding.   Depending on these specific requirements of the algorithm, infrared and collinear safety is not a guarantee.  Selection of the cone size and seed can also cause trigger biases.  However, cone algorithms are very intuitive and are based on the picture that a  jet consists of a large amount of hadronic energy in a small angular region. That is why they have been used as a primary tool to identify jets at hadron colliders starting since the early 1980s.  We propose to use a simple cone algorithm with a seed cut and without the split-merge step to study  and estimate the background subtraction of the complex heavy ion collisions.  

\begin{figure}[ht!]
\begin{center}
$\begin{array}{cc}
\includegraphics[height=6.0cm,clip]{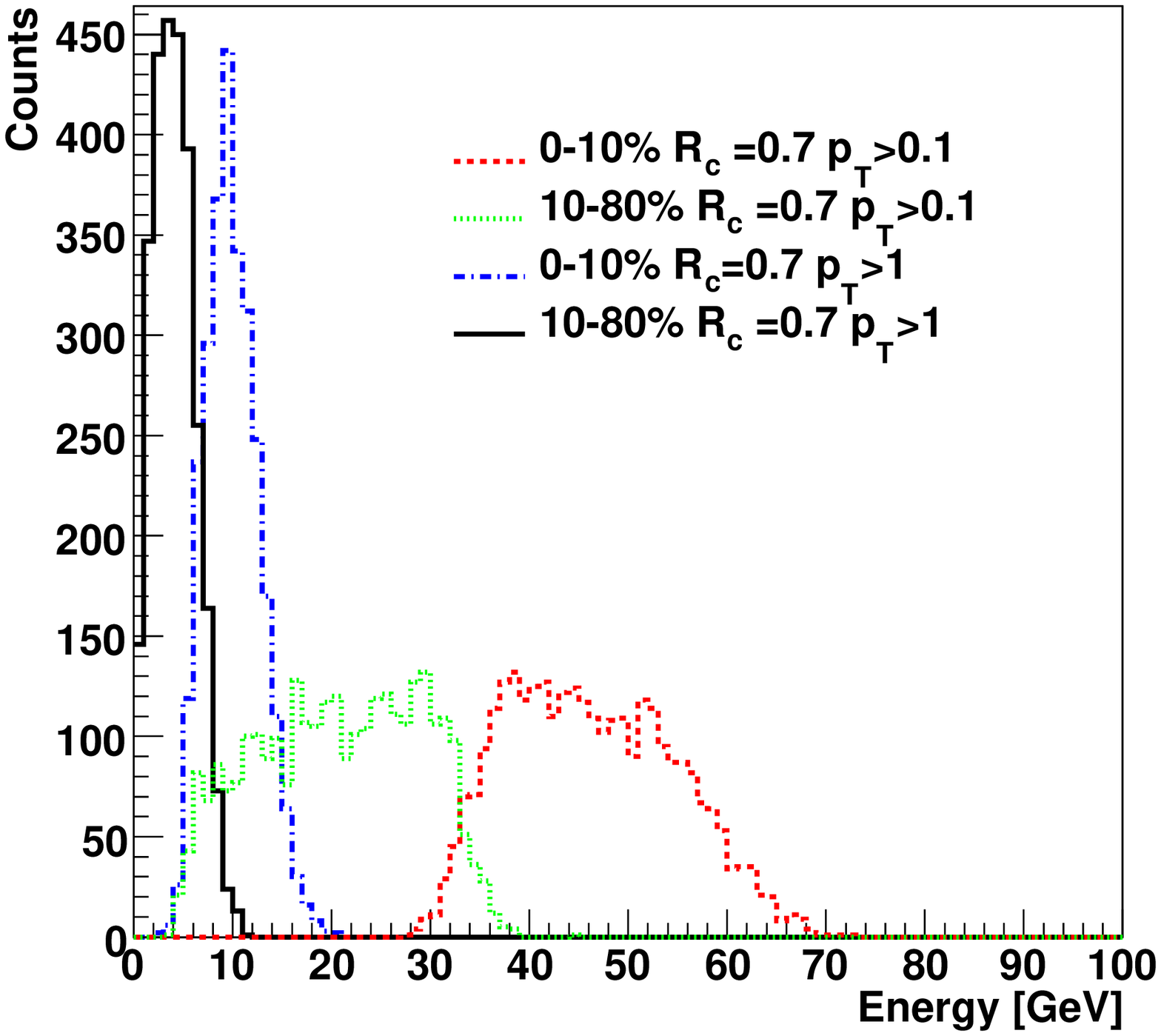} &
\includegraphics[height=6.0cm,clip]{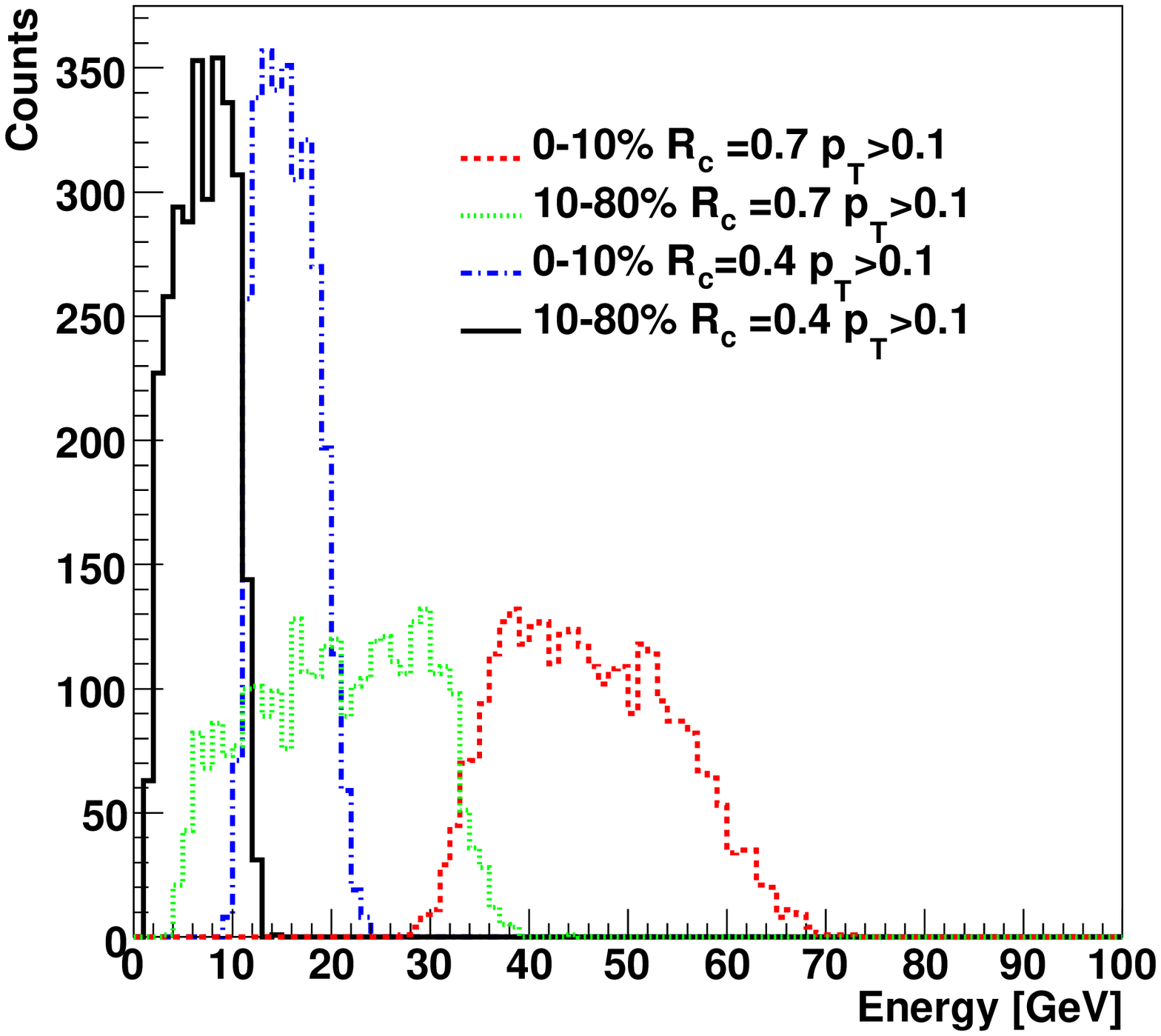} \\
\mbox{\bf (a) }&\mbox{\bf (b)}
\end{array}$

\caption[]{Comparison of the total energy {\bf(a) }for the two $p_{T}$ cuts (0.1 and 1 GeV) for a fixed $R=0.7$ and {\bf(b)} with two cone sizes ($R=0.4$ and $R=0.7$) for a fixed $p_{T}$ cut.  For these plots Hijing events are generated and separated into most central and peripheral Au+Au collisions at $\sqrt{s_{NN}} =200$ GeV. 
}
\label{fig1}
\end{center}
\end{figure}

Figure \ref{fig1}-a shows a comparison of the total energy within a cone size of 0.7 for the two $p_{T}$ cuts for the most central and peripheral $Au+Au$ collisions of simulated Hijing events \cite{hijing}. For larger $p_{T}$ cut ($p_{T}>1$ GeV)  the total energy in a jet cone is reduced significantly  ($\sim 50\%$) with respect to no $p_{T}$ cut, lowering the total background energy collected in a jet cone. 
Figure \ref{fig1}-b shows a similar comparison for the total energy in a jet cone within two different cone sizes when a fixed $p_{T}$ cut is applied. Reducing the cone size also reduces the net energy collection in the jet cone as can be seen as a shift towards the left in the energy distributions.  By using a smaller cone size and a larger $p_{T}$ cut, it is possible to suppress the heavy ion background. The smaller cone size also implies a reduction of the total measured jet energy. However a small cone size of $R=0.4$ can be appropriate, as  $\sim80$\% of the jet energy is observed to be within R$\sim0.3$ for 50 GeV jets in the Tevatron \cite{cdf}.  This assumption is true only if the broadening of the jet fragmentation due to quenching in the medium is small.   Even with the larger $p_{T}$ threshold, additional background subtraction is needed. One possible way is to subtract  the rest of the background from the averaged out-of-jet-cone event-by-event for the selected cone size.  Quantitative studies are needed when selecting regions of subtraction due to $\eta$ and $\phi$ acceptance, $R$ and $p_{T}$ cuts, fluctuations and elliptic flow observed in real heavy ion collisions. The overall energy resolution, detector efficiency and acceptance can be estimated and corrected by simulated  quenched or unquenched fragmentation of jets that are embedded in real heavy ion events.   
 
To study the effect of the background in jet reconstruction for the LHC collisions, quenched jets are generated and reconstructed with a simple cone algorithm without any iterations or split-merge steps in $p+p$ and $Pb+Pb$ collisions  \cite{sarah,joern}.  In these studies, the $Pb+Pb$ background is observed to have a minor effect on the reconstructed jet-energy. It is  also observed that the heavy-ion background is manageable as the differential cross section of reconstructed jet energy with and without $Pb+Pb$ background seems to agree well for jets with  $E_{T}>50$ GeV.

\subsection{Sequential Recombination}\label{kt}

The sequential recombination algorithms, also known as the $k_{T}$  algorithms, have been used extensively in the Tevatron as they are collinear and infrared safe \cite{kt}. In these type of clustering algorithms, arbitrary shaped jets are allowed to follow the energy flow resulting in less bias on the reconstructed jet shape than with 
the seed base cone algorithms \cite{catchment}.    The $k_{T}$ algorithms include various steps such as clustering objects close in relative $p_{T}$, merging these particles into a new cluster when the momentum satisfies a predefined minimum, and repeating steps until all clusters become jets.  Details of this type of algorithm can be found in \cite{cdf,kt} and references therein.   While the steps in combination of particles in sequential recombination algorithms might not be as intuitive as the ones from the cone algorithms, the framework of clustering algorithms such as fastjet  are available and easy to implement on experimental data  \cite{fastjet}.  With the assumption that underlying event and pile-up are distributed uniformly in $\eta$ and $\phi$, an active area ($A_{j}$) of each jet in the fastjet is estimated by filling an event with many very soft particles and then counting how many are clustered into a given jet.  Study of $P_{T}/A_{j}$ of jets determines the noise density and can be subtracted from the measured jet energy on an event-by-event basis to correct for the real jet energy. This correction is observed to recover most of the simulated jet's energy when they are reconstructed in pile up and heavy ion backgrounds \cite{catchment}. A new reverse clustering type called anti-$k_{T}$ is also available within the fastjet algorithm framework \cite{antikt}.
The anti-$k_{T}$ algorithm provides a fast infrared and collinear safe replacement of the traditional cone algorithm and fixes the  irregularities in the boundaries of the final jets due to the soft radiation observed in regular $k_{T}$ type algorithms. Since jet areas can be subtracted as in the $k_{T}$ case, this reverse sequential clustering algorithm is also appropriate for the heavy ion environment and is essential to implement on heavy ion events as a tool to study the systematics of jet reconstruction.

\section{Conclusions}\label{concl}

The LHC is expected to start at the end of the  summer 2008 and will be a jet factory.   At these high energies, the heavy ion background will be larger but also there will be copious production of jets that will be sufficient enough to get above the heavy ion background \cite{peter}.   High cross-section rates of jet production provide sufficient energy lever-arm to map out the energy dependence of jet quenching.  At RHIC the backgrounds are smaller but also cross-sections of high momentum jets are smaller.  Jet structure changes due to energy loss and the additional radiated gluons are expected to be observed at LHC with respect to RHIC.  It is essential to reconstruct jets at LHC and RHIC to measure the energy dependence of the medium induced energy loss to map out all energy levels.  The detector set-ups like calorimeters and time projection chambers are available from both the STAR experiment at the RHIC and the ALICE at the LHC.  The clustering and cone algorithms provide the necessary systematic study of infrared and collinear safe jet reconstruction and of the variation of techniques applied in subtraction of large heavy ion underlying events and fluctuations. While the energy resolution depends on fragmentation functions of hard scattering cross sections, the question of quenching can only be addressed by theoretical input.

\section*{Appendix}\label{app}
First jet reconstruction measurement in most central $Au+Au$ collisions at $\sqrt{s_{NN}}=200$ GeV from Year 2007 run at the RHIC and the systematic study of utilizing both cone and cluster algorithms are presented during the Hard Probes meeting in June 2008 at Illa da Toxa Galicia in Spain \cite{me,jor}.


\vfill\eject

\begin{thebibliography}{9}  
  
\bibitem{drell}S.D Drell, D.J.Levy and T.M. Yan, {\it Phys. Rev.} 187 2159 (1969).
\bibitem{cabibbo}N. Cabibbo, G. Parisi and M. Testa, {\it Lett. Nuovo Cimento} 4,35 (1970).
\bibitem{bjorken}J.D. Bjorken and S.D. Brodsky, {\it Phys. Rev. D} 1, 1416 (1970).
\bibitem{weinberg}Sterman and Weinberg, {\it Phys. Rev. Lett.} 39, 1436 (1977).
\bibitem{cdfall}A. Abulencia et al. [CDF - Run II Collaboration], {\it Phys. Rev. } D 75 092006 (2007)[Erratumibid.
D 75 119901 (2007)] hep-ex/0701051; A. Abulencia et al. [CDF II Collaboration],
{\it Phys. Rev. Lett. }96 122001 (2006) hep-ex/0512062.


\bibitem{cdf}http://www-cdf.fnal.gov/physics/new/qcd/QCD.html. 
\bibitem{starpp}STAR Collaboration, {\it Phys. Rev. Lett.} 97 252001 (2006). 


\bibitem{starpt}STAR Collaboration, {\it Phys. Rev. Lett.} 91 072304 (2003), \\STAR Collaboration, {\it Phys. Rev. Lett.} 97 162301 (2006).
\bibitem{phenixpt}PHENIX Collaboration, {\it Phys. Rev. Lett.} 96 202301 (2006).


\bibitem{borvi}Borghini and Wiedemann, hep-ph/0506218.

\bibitem{eskola}Eskola et al., hep-ph/0406319.
\bibitem{blazey}Gerald C. Blazey et al. FERMILAB-CONF-00-092-E, hep-ex/0005012.
\bibitem{salamtalk}Gavin P. Salam, International Symposium on Multiparticle Dynamics (ISMD07), Berkeley, California, 4-9 Aug 2007, hep-ph 0801.0070. 
\bibitem{jets}Brenna Flaugher and Karlheinz Meier. FERMILAB-CONF-90-248-E, Dec 1990.


\bibitem{pileupcacciari}Matteo Cacciari, Gavin P. Salam {\it Phys.Lett.} B659, 119-126, (2008).

\bibitem{catchment}Matteo Cacciari, Gavin P. Salam, Gregory Soyez, {\it JHEP} 0804:005, (2008), hep-ph/0802.1188. 


\bibitem{grid}For grid system see; http://lcg.web.cern.ch/LCG/

\bibitem{hijing}Miklos Gyulassy, Xin-Nian Wang {\it Comput.Phys.Commun.} 83:307,(1994). 


\bibitem{sarah}S. L. Blyth et al., J. Phys. G 34 271 (2007) ,nucl-ex/0609023.
\bibitem{joern}J. Putschke for the ALICE Collaboration,  Workshop on "Parton fragmentation processes: in the vacuum and in the medium`` Feb 2008
\bibitem{kt}N. Brown, W.James Stirling {\it Z.Phys.} C 53:629-636, (1992).

\bibitem{salam}M. Cacciari, G. Salam hep-ph/0512210.

\bibitem{fastjet}FastJet,M. Cacciari, G. P. Salam and G. Soyez, http://www.lpthe.jussieu.fr/~salam/fastjet.

\bibitem{antikt}Matteo Cacciari, Gavin P. Salam, Gregory Soyez, {\it JHEP} 0804:063,2008, hep-ph/0802.1189.
\bibitem{peter}P.M. Jacobs, M. van Leeuwen, {\it Nucl. Phys.} A774 237-246 (2006).
\bibitem{me}S. Salur for the STAR Collaboration, ``Direct measurement of jets in $\sqrt{s_{NN}}=200$ GeV Heavy Ion Collisions by STAR''
Hard Probes 2008 Proceedings Illa da Toxa (Galicia-Spain). 
\bibitem{jor}J. Putschke for the STAR Collaboration, ``Modified fragmentation measurements with full jet reconstruction in heavy ion collisions at $\sqrt{s_{NN}}=200$ GeV by STAR'' Hard Probes 2008 Proceedings Illa da Toxa (Galicia-Spain).


\end{thebibliography}
\end{document}